\documentclass[smallextended]{svjour3}       
\smartqed  
\usepackage{graphicx}
\usepackage{amsmath,amssymb,bm,mathrsfs,bbold}
\usepackage{bbm}
\usepackage{xspace}
\usepackage{color}
\usepackage[colorlinks=true,linkcolor=blue,urlcolor=blue,citecolor=blue]{hyperref}
\usepackage[normalem]{ulem}
\usepackage{hyperref}

\renewcommand{\S}{{\mathbf S}}
\newcommand{\Stot}{{\mathbf S}_{\rm tot}}

\newcommand{\n}{\mathbf n}

\newcommand{\K}{\mathbf K}
\newcommand{\B}{\mathbf B}

\renewcommand{\a}{\mathbf a}
\renewcommand{\b}{\mathbf b}

\newcommand{\be}{\begin{equation}}
\newcommand{\ee}{\end{equation}}

\begin{document}


\title{A map between time-dependent \\ and time-independent quantum many-body Hamiltonians
}


\author{Oleksandr Gamayun        \and
        Oleg Lychkovskiy 
}


\institute{O. Gamayun \at
              Institute for Theoretical Physics and Delta Institute for Theoretical Physics,\\
University of Amsterdam, Science Park 904, 1098 XH Amsterdam, The Netherlands
           \and
           O. Lychkovskiy \at
             Skolkovo Institute of Science and Technology,\\
Skolkovo Innovation Center 3, Moscow  143026, Russia,\\
and Department of Mathematical Methods for Quantum Technologies, \\ Steklov Mathematical Institute of Russian Academy of Sciences,\\
Gubkina str. 8, Moscow 119991, Russia,\\
and Laboratory for the Physics of Complex Quantum Systems, \\ Moscow Institute of Physics and Technology, \\ Institutsky per. 9, Dolgoprudny, Moscow  region,  141700, Russia\\
\email{o.lychkovskiy@skoltech.ru}
}

\date{Received: July 25, 2020 / Accepted: }

\maketitle

\begin{abstract}
Given a time-independent Hamiltonian $\widetilde H$, one can construct a time-dependent Hamiltonian $H_t$  by means of the gauge transformation $H_t=U_t \widetilde H \, U^\dagger_t-i\, U_t\,  \partial_t U_t^\dagger$. Here $U_t$ is the unitary transformation that relates the solutions of the corresponding Schrodinger equations. In the many-body case one is usually interested in Hamiltonians with few-body (often, at most two-body) interactions. We refer to such Hamiltonians as {\it physical}. We formulate sufficient conditions on $U_t$ ensuring that $H_t$ is physical as long as $\widetilde H$ is physical (and vice versa). This way we obtain a general method for finding such pairs of physical Hamiltonians $H_t$, $\widetilde H$ that the driven many-body dynamics governed by $H_t$ can be reduced  to the quench dynamics due to the time-independent $\widetilde H$. We apply this method to a number of many-body systems. First we review the mapping of a spin system with isotropic Heisenberg interaction and arbitrary time-dependent magnetic field to the time-independent system without a magnetic field [F. Yan, L. Yang, B. Li, Phys. Lett. A 251, 289 (1999);  Phys. Lett. A 259, 207 (1999)]. Then we demonstrate that essentially the same gauge transformation  eliminates an  arbitrary time-dependent magnetic field from a system of interacting fermions. Further, we apply the method to the quantum Ising spin system and a spin coupled to a bosonic environment. We also discuss a more general situation where $\widetilde H = \widetilde H_t$ is time-dependent but dynamically integrable.

\end{abstract}

\section{Introduction}
\label{intro}

Quantum many-body systems driven by external classical parameters varying in time are ubiquitous in nature and in labs. In particular, time-dependent electric and magnetic fields are routinely used to  engineer and manipulate large quantum systems consisting of cold atoms in optical lattices, ions in ion traps, arrays of superconducting qubits and quantum dots {\it etc}.

From the theoretical standpoint, the description of a driven quantum many-body system is, in general, a formidable task.  The state of such system is described by a multidimensional dynamical wave function $\Psi_t$ satisfying the Schr\"odinger equation
\be\label{Schrodinger equation}
i\partial_t \Psi_t = H_t\,\Psi_t,
\ee
where  $H_t$ is the system's time-dependent Hamiltonian. A major difficulty of this equation is that it can not be, in general, reduced to the stationary Schr\"odinger equation, in contrast to the case of a Hamiltonian independent on time. This difficulty adds to the complexity due to the many-body nature of the problem.

In recent years it has been realized that certain  quantum many-body models with driving are {\it dynamically integrable}, i.e. allow for explicit solutions. These are mostly time-dependent generalization of models solvable by Bethe ansatz~\cite{Barmettler_2013,Fioretto_2014,gritsev2017integrable,sinitsyn2018integrable,ermakov2019time} and continuous-variable models with scaling invariance \cite{Kagan_1996_evolution,castin_1996_bose-einstein,castin_2004_exact,minguzzi2005exact,Gritsev_2010,delcampo2013shortcuts,deffner2014classical}.

Here we elaborate upon a somewhat different approach to the driven many-body dynamics. Namely, we consider time-dependent many-body Hamiltonians that can be reduced either to time-independent Hamiltonians (integrable or not) or to dynamically integrable time-dependent Hamiltonians with the help of a suitable gauge transformation. This transformation implies a unitary map between the corresponding dynamical wave functions, and thus the dynamics induced by one Hamiltonian is straightforwardly restored from the dynamics induced by another one. This approach has been previously applied to a number of particular many-body systems \cite{yan1999formal,yan1999invariant,Colcelli2019integrable,gamayun2020nonequilibrium}.

Importantly, we impose a physical requirement that both Hamiltonians contain only few-body interactions, and put forward a class of transformations that automatically satisfy  this requirement. This general construction is described in the next section. In sections \ref{sec:3} -- \ref{sec:6} we apply this machinery to particular systems. A brief summary is given in section \ref{sec:7}. Some technical details are relegated to the Appendices.



\section{Gauge transformation preserving few-body nature of interactions}
\label{sec:2}

\paragraph{Gauge transformation.}~~Consider two time-dependent Hamiltonians, $H_t$ and $\widetilde H_t$. The corresponding wave functions, $\Psi_t$ and $\widetilde \Psi_t$, satisfy the Schrodinger equations \eqref{Schrodinger equation} and
\be\label{Schrodinger equation tilde}
i\partial_t \widetilde\Psi_t = \widetilde H_t\,\widetilde \Psi_t,
\ee
respectively. Obviously, there always exists a unitary transformation $U_t$ with the property $U_0=\mathbb 1$, such that
\be\label{transformation of wave functions}
\Psi_t = U_t\, \widetilde \Psi_t
\ee
for arbitrary common initial conditions $\Psi_0=\widetilde \Psi_0$. It is easy to verify that this unitary transformation of the wave function induces a gauge  transformation of the Hamiltonian,
\be\label{gauge transformation}
H_t=U_t \widetilde H_t \, U^\dagger_t- W_t ,\qquad  W_t\equiv i\, U_t\,  \partial_t U_t^\dagger.
\ee

The above formula is well-known and emerges in many contexts, including quantum electrodynamics \cite{feynman2018quantum}, dynamical symmetries \cite{wang1993algebraic}, adiabatic theorem \cite{kato1950,il'in2020adiabatic},  counterdiabatic driving \cite{demirplak2003adiabatic,demirplak2005assisted,Berry_2009,Kolodrubetz2017,Sels_2017_minimizing} and  Floquet engineering \cite{eckardt2010frustrated,jotzu2014experimental,eckardt2017colloquium} (see also its analog for Lindblad dynamics of open systems  in \cite{Scopa_2018_Lindblad,Scopa_2019_exact}).

Eq. \eqref{gauge transformation} generalizes the transformation to a uniformly rotating reference frame in the Hilbert space given by
\be\label{uniform rotation}
H_t=e^{iWt} \widetilde H e^{-iWt} -W=e^{iWt}\left( \widetilde H  -W\right) e^{-iWt}
\ee
with a time-independent $W$ and $\widetilde H$. The latter transformation has numerous applications, from  Rabi oscillation \cite{Rabi1954} to the electron transport through a driven conformal quantum point contact \cite{gamayun2020nonequilibrium}. An important difference between  eqs. \eqref{uniform rotation} and \eqref{gauge transformation} is that $H_t$ in eq. \eqref{uniform rotation}  has a constant spectrum, while  $H_t$ in eq.~\eqref{gauge transformation} can have a spectrum varying in time in an arbitrary way.

\paragraph{Preserving few-body nature of interactions.}~~Since the gauge transformation~\eqref{gauge transformation} can relate two  arbitrary time-dependent Hamiltonians acting on the same Hilbert space, it is tempting to use this transformation along with eq.~\eqref{transformation of wave functions} to reduce the dynamics governed by some complex time-dependent $H_t$ to the dynamics governed by some simple, preferably time-independent $\widetilde H_t$. One can also attempt to turn the tables and generate a new tractable Hamiltonian  $H_t$ from a known tractable (e.g. time-independent or dynamically integrable) Hamiltonian $\widetilde H_t$.

However, when it comes to the many-body systems, the set of tractable and/or physically meaningful Hamiltonians is normally restricted by the Hamiltonians with few-body (often, at most two-body) interactions. We will refer to such Hamiltonians as {\it physical} throughout the paper. It is then natural to ask under what conditions both $H_t$ and $\widetilde H_t$ in eq. \eqref{gauge transformation} are physical in this sense. Here we come to the central observation of the present paper: To ensure that both  $H_t$ and $\widetilde H_t$ are physical it is sufficient to require that
\be\label{factorized U}
U_t=\prod_{j=1}^K U_{j}(t),
\ee
where each unitary $U_j(t)$ is a few-body operator, $K$ is at most polynomial in the system size $L$ and
\be\label{commutator}
 [U_j(t), U_{j'}(t')]=0\qquad {\rm for} \qquad j\neq j'.
\ee
Indeed, under these conditions eq.~\eqref{gauge transformation} implies that whenever $\widetilde H_t$ is an extensive sum of few-body terms, so is  $H_t$.
The commutation relation \eqref{commutator} at $t=t'$ (at $t\neq t'$) ensures the few-body nature of $U_t \widetilde H U_t^\dagger$ (of $W_t$). It should be emphasized that commutation of $U_j(t)$  with $U_j(t')$ is not required.

In general, there is no guarantee that there exists a unitary transformation relating two given physical Hamiltonians according to eq. \eqref{gauge transformation} and satisfying the conditions \eqref{factorized U}, \eqref{commutator}. However, we will demonstrate that such unitary transformation exists for a number of important  models.

\paragraph{Recipe summary.}~~Let us summarize our main idea. Assume one wishes to study quantum many-body dynamics governed by a physical time-dependent many-body Hamiltonian $H_t$ but is unable to solve the Schrodinger equation~\eqref{Schrodinger equation} directly. One can then attempt to find a unitary operator $U_t$ of the product form \eqref{factorized U} satisfying the condition~\eqref{commutator}, such that the gauge transformation  \eqref{gauge transformation} reduces $H_t$ to a more tractable $\widetilde H_t$. The conditions \eqref{factorized U}, \eqref{commutator} automatically ensure that $\widetilde H_t$ is also physical. If one is then able to solve the Schrodinger equation~\eqref{Schrodinger equation tilde} with $\widetilde H_t$, then the solution to the original Schrodinger equation~\eqref{Schrodinger equation} is obtained by the unitary transformation~\eqref{transformation of wave functions}.

While this recipe may seem to be quite abstract at this point, we will substantiate it by considering specific models in what follows. Before we turn to these models in the next sections, let us present some additional general considerations and introduce  relevant notions.

\paragraph{Special initial states.}~~Assume $\widetilde H_t=\widetilde H$ is independent on time, and the system is initialized in an eigenstate of $\widetilde H$ with the eigenvalue $\widetilde E$. Then
\be\label{special dynamics}
\Psi_t=e^{-i\widetilde E t} \, U_t\, \Psi_0.
\ee
This implies that the dynamics of any few-body observable~$A$ can be found explicitly, $\langle \Psi_t |A| \Psi_t \rangle= \langle \Psi_0 |U_t^\dagger A \, U_t| \Psi_0 \rangle$, provided the few-body expectation values of the initial state are known. This follows from the fact that $U_t^\dagger A \, U_t$ is a few-body operator thanks to eqs. \eqref{factorized U},\eqref{commutator}. We give an explicit example of the dynamics of a special initial state in section \ref{sec:3}.

\paragraph{Floquet dynamics.}~~Assume that $\widetilde H_t=\widetilde H$ is independent on time and $U_t$ is periodic with the period $T$. This implies that $H_t$ is also periodic, and we immediately obtain the stroboscopic Floquet dynamics
\be\label{special dynamics}
\Psi_{n T}=e^{-i n \widetilde H T}\,  \Psi_0, \quad n\in \mathbb{Z},
\ee
with the Floquet Hamiltonian equal to $ \widetilde H$.

Remarkably, thus obtained exact Floquet Hamiltonian  contains only few-body interactions.   This is in sharp contrast to typical Floquet Hamiltonians involving arbitrarily extended many-body terms \cite{eckardt2017colloquium}. One particular consequence of this feature is that the periodic driving fails to  heat the system to the infinite temperature, contrary to what is happening in generic systems~\cite{dalessio2014long-time,lazarides2014equilibrium,ponte2015periodically}. It has been actually anticipated that if the time dependence can be eliminated by moving to a different reference frame, the indefinite heating hypothesis will fail \cite{dalessio2014long-time}. Here we provide a receipt how to construct a broad range of local many-body models where this indeed happens.

\paragraph{Integrable driven dynamics.}~~If  $\widetilde H_t=\widetilde H$ is independent on time and integrable, the driven dynamics of $H_t$ is reduced to the integrable quench dynamics of $\widetilde H$.  Remarkably, the instantaneous integrability of $H_t$ can be absent in this case.

Integrals of motion $\widetilde I_n$ of $\widetilde H$ are mapped to the {\it dynamical invariants}
\be\label{dynamical invariant}
I_n(t)= U_t \tilde I_n U_t^\dagger.
\ee
Dynamical invariants are operators that are, in general, explicitly time-dependent, satisfy the equation $i\partial_t I_n(t)=[H_t,I_n(t)]$ and thus possess conserved expectation values, $\partial_t \,\langle \Psi_t| I_n(t) | \Psi_t\rangle=0$ \cite{lewis1969exact}.

Further, if $\widetilde H_t$ is time-dependent and dynamically integrable, so is $H_t$. This statement is illustrated by an explicit example in section   \ref{sec:5}.

\paragraph{Variational approach and relation to counterdiabatic driving.}~~ While a gauge transformation is a standard tool in constructing counderdiabatic driving protocols \cite{demirplak2003adiabatic,demirplak2005assisted,Berry_2009,Kolodrubetz2017,Sels_2017_minimizing}, the meaning of $U_t$ there is very different from that in the present paper. Namely, here $U_t$ relates solutions of the {\it dynamical} Schrodinger equations with different Hamiltonians, while there   $U_t$ relates instantaneous eigenstates of a time-dependent Hamiltonian at different times. As a consequence, $W_t$ in eq. \eqref{gauge transformation} is {\it not} the adiabatic gauge potential introduced in \cite{Kolodrubetz2017}.

Nevertheless, formal similarities are strong enough to attempt adapting tools developed for the counteradiabatic driving to the present problem. In particular, one can observe, following ref. \cite{Sels_2017_minimizing}, that eq. \eqref{gauge transformation} with time-independent  $\widetilde H_t=\widetilde H$  entails\footnote{We are grateful to A. Polkovnikov for pointing to this equation.}
\be\label{equation for Wt}
\partial_t H_t=i[W_t,H_t]- \partial_t W_t.
\ee
Note that this equation does not contain $\widetilde H$. Given $H_t$, one can try to find $W_t$ from this equation. If successful, one can then restore $U_t$ from $W_t$ and find $\widetilde H$ from eq. \eqref{gauge transformation}. To summarize, solving eq. \eqref{equation for Wt} is equivalent to finding $U_t$ and $\widetilde H$ for a given $H_t$. One can also address eq. \eqref{equation for Wt}  variationally, as in ref. \cite{Sels_2017_minimizing}. We leave further research in this promising direction for future work.


\section{Heisenberg model with time-dependent magnetic field}
\label{sec:3}

\paragraph{Model.}~~Here we consider a system of quantum spins on an arbitrary lattice, with the isotropic Heisenberg interaction and a time-dependent, spatially homogeneous magnetic field $\B_t$:
\be\label{Heisenberg with Bt}
H_t=H_{\rm H}-\B_t \Stot,
\ee
where  $ \Stot=\sum_j \S_j$ is the total spin and
\be\label{Heisenberg}
H_{\rm H}= \sum_{i<j}  J_{ij}\, {\bf S_i}  {\bf S_j}
\ee
describes the isotropic Heisenberg coupling between the spins with arbitrary coupling constants $J_{ij}$, indices  $i,j$ labeling the sites of the lattice. We emphasise that our reasoning will apply to spins with arbitrary spin quantum number (not necessarily spins $1/2$).

The fact that the Hamiltonian \eqref{Heisenberg with Bt} can be essentially reduced to a time-independent Hamiltonian was arguably first recognized in refs. \cite{yan1999formal,yan1999invariant}. We will first present this mapping in our own framework, and then comment on the presentation of refs. \cite{yan1999formal,yan1999invariant}.

\paragraph{Eliminating magnetic field.}~~It turns out that for an arbitrary $\B_t$ one can choose such $U_t$ of the form~\eqref{factorized U},\eqref{commutator} that the time-dependent Hamiltonian \eqref{Heisenberg with Bt} can be mapped onto the time-independent Hamiltonian $\widetilde H=H_{\rm H}$ by the gauge transformation \eqref{gauge transformation}. This $U_t$ reads
\be\label{U for Heisenberg}
U_t=\exp\left( i \,{\bf  K}_t  \Stot \right),
\ee
where ${\bf  K}_t$ is the solution of the system of differential equations
\begin{align}\label{resolved}
\dot K_t =& \B_t \n_t,  \nonumber \\
\dot \n_t= & \frac12\, \n_t\times \B_t + \frac12 \cot \frac{K_t}2 \,\big( \B_t -(\B_t \n_t) \, \n_t \big),
\end{align}
with $\n_t$ being a unit vector and  ${\bf  K}_t=K_t \n_t$.
Here and in what follows the dot stands for  the time derivative and the cross ``$\times$''stands for the vector product. The initial condition is ${\bf  K}_t=0$ which ensures $U_0=\mathbb 1$. The ambiguity in the initial condition for $\n_t$ is convenient to resolve as $\n_0 =\B_0/|B_0|$.

The derivation of the system of equations \eqref{resolved}, as well as its alternative forms, are presented in the Appendix \ref{appendix A}.  The key point of the derivation is that $U_t H_{\rm H} U_t^\dagger=H_{\rm H}$ due to the rotational invariance of the isotropic Heisenberg interaction.

Interestingly,  $U_t$ can be expressed in a quite different form tagged as ``Gauss parametrization'' \cite{ringel2013dynamical,wei1963lie}. A remarkable feature of the Gauss parametrization is that it leads to a Riccati equation. We review this result in Appendix \ref{appendix B}.

Let us briefly comment on refs. \cite{yan1999formal,yan1999invariant} and related work. In \cite{yan1999formal} the Hamiltonian \eqref{Heisenberg with Bt} was reduced to a Hamiltonian with Heisenberg interactions and magnetic field with a time-dependent magnitude but fixed direction along the $z$ axis. This was done by means of a unitary transformation similar to that in eq.~\eqref{Gauss unitary transformation}. This  transformation had been  earlier used to treat a time-dependent Hamiltonian constructed of a sum of  SU(2) and SU(1,1) generators~\cite{lai1996time-dependent}  (in particular, a spin in a time-dependent magnetic field could be treated by the method of \cite{lai1996time-dependent}). In \cite{yan1999invariant} the  Hamiltonian \eqref{Heisenberg with Bt} was reduced to a time-independent one in a somewhat indirect way in the framework of the theory of dynamical invariants \cite{lewis1969exact}. A dynamical invariant for a single spin in a time-dependent magnetic field was constructed in~\cite{Gao1991formally}.

We also remark that since  $H_t$ and $H_{t'}$ form a finite closed algebra with respect to commutation, the dynamics can be alternatively addressed by a technique described in \cite{gritsev2017integrable}.

We note that apart from the time-dependent dynamical invariant constructed in \cite{Gao1991formally,yan1999invariant}, the Hamiltonian \eqref{Heisenberg with Bt} has two independent conserved quantities, $H_{\rm I}$ and $\Stot^2$. Further, if $H_{\rm I}$  is one-dimensional with  nearest-neighbour Heisenberg interactions (i.e. integrable), there is a large number of additional local, rotationally-invariant,  time-independent conserved quantities that can be found e.g. in ref. \cite{Grabowski1994quantum}.

\paragraph{Special initial states.}~~ It follows from eq. \eqref{special dynamics} and the above considerations that any  eigenstate of $\widetilde H =H_{\rm H}$ evolves as if there were no interactions between spins whatsoever, i.e. under the action of the Hamiltonian $(-\B_t \Stot)$  (up to an irrelevant overall phase). The most simple eigenstates of  $H_{\rm H}$ are product states of spins pointing in the same direction, and this form of the state vector is preserved throughout the evolution, the direction of a spin being determined by the Schr\"odinger equation with a single-particle Hamiltonian $(-\B_t \S_i)$.

\section{Interacting fermions in a time-dependent magnetic field}
\label{sec:4}

Here we consider a general Hamiltonian of interacting lattice fermions with spin $1/2$ in an external homogeneous time-dependent magnetic field,
\be\label{interacting fermions in magnetic field}
H_t=H_e - \sum_{\alpha,\sigma,\tilde\sigma} B_t^\alpha  s_\alpha^{ \tilde\sigma \sigma} \sum_{i} c^\dagger_{\tilde\sigma i} c_{\sigma i},
\ee
where
\be\label{interacting fermions}
H_e=-\frac12 \sum_{\sigma, i,j} \epsilon_{ij} c^\dagger_{\sigma i} c_{\sigma j} + \frac12 \sum_{\sigma, i,j} V_{ij} c^\dagger_{\sigma i} c_{\sigma i} c^\dagger_{\sigma j} c_{\sigma j},
\ee
$i,j$ label lattice sites, $\sigma, \tilde \sigma\in \{\uparrow,\downarrow\}$ are spin indices, $\alpha\in\{x,y,z\}$ labels $2\times2$ spin matrices $s_\alpha$ equal to  the corresponding Pauli matrices divided by~2, $c_i$ is a fermion annihilation operator at the lattice site $i$, $\epsilon_{ij}$ are tunneling matrix elements (for $i\neq j$) or local on-site potentials (for $i=j$), $V_{ij}$ is a two-body interaction potential and $\B_t$ is the time-dependent external magnetic field. A particular case of this Hamiltonian is the Hubbard model, that is integrable in one dimension \cite{essler2005one}.

In complete analogy with the argument in the previous section, we find that an arbitrary $\B_t$ can be eliminated by the gauge transformation \eqref{gauge transformation} with $\widetilde H=H_{e}$ and
\be
U_t=\exp\Big( \sum_{\alpha,\sigma,\tilde\sigma} K^\alpha_t s_\alpha ^{ \tilde\sigma \sigma} \sum_{i} c^\dagger_{\tilde\sigma i} c_{\sigma i} \Big),
\ee
where $\K_t$ satisfies differential equations \eqref{resolved}.

Alternatively, one can use $U_t$ in the Gauss parametrization, analogously to eq. \eqref{Gauss unitary transformation}.

Note that the results of the present section can be straightforwardly generalized to fermions or bosons with an arbitrary spin, to models with continuous variables and  to Kondo-type models with both itinerant and localized particle species.

\section{Ising model in a time-dependent magnetic field}
\label{sec:5}

\paragraph{Ising model on a general lattice.}~~Here we consider the quantum Ising model in a time-dependent magnetic field,
\be\label{Ising with Bt}
H_t=H_{\rm I}-\sum_i \B_{i}(t) \S_i,
\ee
where
\be\label{Heisenberg}
H_{\rm I}= \sum_{i<j}  J_{ij}\,  S_i^x   S_j^x
\ee
describes the Ising coupling. As in Sect. \ref{sec:3}, the spins reside on an arbitrary lattice labeled by indices $i$, $j$.
In contrast to the previous examples, now the magnetic field $\B_{i}(t)$ need not be homogeneous, which is indicated by the index~$i$. For the latter reason we have to use a somewhat different notation for the magnetic field, $\B_{i}(t)$ instead of $\B_t$. In what follows the argument $t$ in $\B_{i}(t)$ will be sometimes omitted for brevity.

We introduce  $U_t$ of the form \eqref{factorized U}, where each individual $U_i(t)$ reads
\be\label{U for Ising}
U_i(t)=\exp(i\, \phi_i(t) S^x_i)
\ee
and phases $\phi_i(t)$ are defined as
\be\label{phi}
\phi_i(t)=-\int_0^t B^x_i(t') \,dt'.
\ee
Then, provided that the magnetic field satisfies
\begin{align}\label{BxBy}
B^y_i(t)= & B^y_i(0)\, \cos \phi_i(t) + B^z_i(0)\, \sin \phi_i(t), \nonumber \\
B^z_i(t)= & -B^y_i(0) \, \sin \phi_i(t) +B^z_i(0)\, \cos \phi_i(t),
\end{align}
the time-dependent Hamiltonian \eqref{Ising with Bt} can be mapped to the time-independent Hamiltonian
\be\label{tilde H Ising}
\widetilde H=H_{\rm I} -\sum_i \left( B_{i}^y(0) \, S_i^y+ B_{i}^z(0) \, S_i^z\right)
\ee
by means of the gauge transformation \eqref{gauge transformation}.

Note that, in contrast to the previous cases, here we are able to eliminate the time dependence of the magnetic field  only for such $\B_i(t)$ that satisfy eqs. \eqref{BxBy}, \eqref{phi}.

Also note that, in contrast to previous cases,  $H_t$ and $H_{t'}$ here do not form a finite closed algebra with respect to commutation, and the technique of ref.~\cite{gritsev2017integrable} can not be utilized.

Remarkably, in the case of one-dimensional nearest-neighbour model the Hamiltonian \eqref{tilde H Ising} is integrable \cite{Pfeuty1970}. This motivates us to consider the latter case separately.


\paragraph{Nearest-neighbour one-dimensional Ising model.}~~The Hamiltonian of this model reads
\be\label{1D Ising with Bt}
H_t= \sum_i  J_i\,  S_i^x   S_{i+1}^x-\sum_i \B_{i}(t) \S_i.
\ee

Assume that the magnetic field satisfies
\begin{align}\label{BxBy integrable}
B^y_i(t)= & B^\bot_i(t) \Big( B^y_i(0)\, \cos \phi_i(t) + B^z_i(0)\, \sin \phi_i(t) \Big), \nonumber \\
B^z_i(t)= & B^\bot_i(t) \Big( -B^y_i(0) \, \sin \phi_i(t) +B^z_i(0)\, \cos \phi_i(t) \Big),
\end{align}
where
\be
B^\bot_i(t)=\sqrt{(B^y_i)^2 + (B^z_i)^2}
\ee
and $\phi_t$ is given by eq. \eqref{phi}. It should be emphasized that these conditions are less restrictive than the conditions \eqref{BxBy}. In particular, they allow for a {\it time-dependent} magnitude $B^\bot_i(t)$ of the transverse component of the magnetic field.

Under these conditions the gauge transformation \eqref{gauge transformation} reduces the Hamiltonian \eqref{BxBy integrable}  to the Hamiltonian
\be\label{tilde H Ising integrable}
\widetilde H_t=H_{\rm I} -\sum_i B^\bot_i(t) \Big( B_{i}^x(0) \, S_i^x+ B_{i}^y(0) \, S_i^y\Big),
\ee
where the unitary transformation $U_t$ is given by eq. \eqref{factorized U} with $U_i(t)$ defined in eq.~\eqref{U for Ising}.

In contrast to previous examples, here  $\widetilde H_t$ is time-dependent. However, this Hamiltonian is still tractable, since it can be mapped to  a {\it quadratic} fermionic Hamiltonian with time-dependent coefficients by means of the Jordan-Wigner transformation \cite{Pfeuty1970,lieb1961two,Dziarmaga2005,das2010exotic}. This way we are able to reduce the time-dependent {\it many-body} problem \eqref{1D Ising with Bt} to the  {\it one-body} problem of a single fermion in a time-dependent lattice.

\paragraph{Comment on instantaneous (non-)integrability.}~~Remarkably, the magnetic field in the Hamiltonian  \eqref{1D Ising with Bt} has,  in general, a non-zero longitudinal component $B_i^x(t)$. Consequently, this Hamiltonian is  dynamically integrable despite being nonintegrable at a fixed $t$. This contrasts to refs. ~\cite{Barmettler_2013,Fioretto_2014,gritsev2017integrable,sinitsyn2018integrable,ermakov2019time} where dynamical integrability is shown for Hamiltonians that are instantaneously integrable at any moment of time.

\paragraph{Comment on localization.}~~Assume that  the magnetic field in $\widetilde H$ given by eq. \eqref{tilde H Ising} is homogeneous, $B_{i}^y(0) = B^y(0),~  B_{i}^z(0)=B^z(0)$, and $H_{\rm I}$ does not contain any disorder, e.g.  it is translation-invariant. Curiously,  one still has a freedom to construct $H_t$ that is many-body localized at any fixed~$t$  by choosing appropriate disordered phases $\phi_i(t)$ in the unitary transformation \eqref{U for Ising}. The lesson that we learn from this example is that  even if some $H_t$ is many-body localized at any fixed $t$, its dynamics may be reduced to a quench dynamics of a Hamiltonian without disorder and localization.

\paragraph{Comment on driving through a critical point.}~~In ref. \cite{Dziarmaga2005} the Ising model driven through a quantum critical point is studied and the Kibble-Zurek mechanism for the density of the created defects is verified. While  a particular driving protocol is considered in ref. \cite{Dziarmaga2005},  our method allows one to extend their analysis to a much wider range of paths in the parameter space crossing a phase transition. Note, however, that our method does not produce a counterdiabatic (or  transitionless) driving, as in refs. \cite{delcampo2012assisted,Saberi_2014_adiabatic,Sels_2017_minimizing}, see the discussion in Sect.~\ref{sec:2}.

\section{Spin coupled to bosonic environment}\label{sec:6}

Here we consider a spin in a time-dependent magnetic field coupled to a bosonic environment, the total Hamiltonian being
\be\label{spin-boson}
H_t=-\B_t \, \S +   S^x\sum_k f_k (a^\dagger_k+a_k)+\sum_k \omega_k\, a_k^\dagger a_k.
\ee
Here $a_k$ is the operator annihilating the bosonic mode with the quantum number $k$, $\omega_k$ is the energy of the corresponding mode and $f_k$ is the spin-boson coupling (it usually scales with the system size $L$ and $1/\sqrt{L}$). In the case of spin $1/2$ this Hamiltonian can describe a qubit coupled to bosonic modes.

In complete analogy with the Ising model, we find that the Hamiltonian $H_t$ given by eq. \eqref{spin-boson} can be reduced to the time-independent \be\label{tilde H spin-boson}
\widetilde H=-\left( B^y_0 \, S_i^y+ B^z_0 \, S_i^z\right) +   S^x\sum_k f_k (a^\dagger_k+a_k)+\sum_k \omega_k\, a_k^\dagger a_k
\ee
by means of the gauge transformation \eqref{gauge transformation} with $U_t$ given by eq. \eqref{U for Ising}, provided the magnetic field satisfies the conditions \eqref{BxBy} (where indices $i$ in spin and magnetic field variables should be omitted).

\section{Summary}\label{sec:7}
To summarize, we have presented a method to map certain time-dependent many-body Hamiltonians to time-independent Hamiltonians or to more simple (e.g. dynamically integrable) time-dependent Hamiltonians while preserving the few-body nature of interactions. The method is based on a gauge transformation \eqref{gauge transformation} with a unitary operator of a product form \eqref{factorized U} satisfying the condition \eqref{commutator}.  We have applied this general method to eliminate time-dependent magnetic fields from Heisenberg and Ising quantum spin systems, from a system of a spin coupled to a bosonic environment, as well as  from a system of interacting fermions. The method opens new perspectives to study dynamical integrability, Floquet dynamics of periodically driven systems and driving through quantum critical points.

\begin{acknowledgements} We thank Vladimir Gritsev for useful discussions. We are also grateful to Adolfo del Campo, Anatoli Polkovnikov and Stefano Scopa for valuable remarks.
The work of O.L. was supported by the Russian Foundation for Basic Research under the grant No 18-32-20218.
\end{acknowledgements}

\begin{appendix}
\section{Eliminating magnetic field: covariant parametrization \label{appendix A}
}

Here we outline the derivation of eq. \eqref{resolved}. First we use a formula for the derivative of the exponential map \cite{blanes2009magnus} to obtain an integral representation of  $W_t$:
\begin{align}\label{derivative of exponent magnetic field}
W_t=i e^{i\K_t \Stot}\frac{d}{dt} e^{-i\K_t \Stot} = \int_0^1 dx\,  e^{i x \,\K_t \Stot} \, \left(\dot \K_t \Stot\right)e^{-i x\, \K_t \Stot}.
\end{align}
To proceed further, we use
\begin{align}\label{bch}
e^{i\, \a \S} \, (\b\S) \, e^{-i \,\a \S}&=   \frac1{a^2}  (\a \b) (\a \S)-  \frac1{a} (\a \b \S)  \sin a \nonumber \\
& +\left( (\b \S)- \frac1{a^2}  (\a \b) (\a \S) \right) \cos a
\end{align}
valid for arbitrary vectors $\a$ and $\b$ and arbitrary spin $\S$.  Here $(\a\b)$ denotes the scalar product and $(\a \b \S)$ denotes the scalar triple product. With the help of this formula eq. \eqref{derivative of exponent magnetic field} can be explicitly integrated. Taking into account that $U_t H_{\rm H} U_t^\dagger=H_{\rm H}$ and, consequently, $\B_t \Stot =W_t $ according to eqs. \eqref{gauge transformation},\eqref{Heisenberg with Bt},\eqref{Heisenberg},\eqref{U for Heisenberg}, we obtain the equation
\begin{align}\label{differential equation}
\B_t = &\,\frac{\sin K}{K} \dot\K_t -\frac{1-\cos K}{K^2} \K_t\times\dot \K_t \nonumber\\
       & \,+\frac1{K^2}\left(1-\frac{\sin K}{K}\right)\left(\K_t\dot\K_t\right) \K_t
\end{align}
Introducing $\K_t=K_t \n_t$, where $\n_t$ is a unit vector, one reduces eq. \eqref{differential equation} to
\begin{align}\label{differential equation covariant}
\B_t = \dot K_t\, \n_t+\sin K_t \,\,\dot \n_t -(1-\cos K_t) \,(\n_t\times\dot\n_t).
\end{align}
By performing a scalar (vector) multiplication of this equation by  $\n_t$ (and doing some additional algebra in the second case), one obtains the first (the second) line of eq. \eqref{resolved}. Note that eqs. \eqref{differential equation} and \eqref{differential equation covariant} can be more suitable for numerical integration than eq. \eqref{resolved}.

\section{Eliminating magnetic field: Gauss parametrization \label{appendix B} }

Gauss parametrization of $U_t$ reads \cite{ringel2013dynamical}
\be\label{Gauss unitary transformation}
U_t=\exp\left(   \xi^+_t S_{\rm tot}^+\right) \exp\left(  \xi^z_t S_{\rm tot}^z \right) \exp\left(  \xi^-_t S_{\rm tot}^-\right),
\ee
where $S_{\rm tot}^\pm=S_{\rm tot}^x \pm i S_{\rm tot}^y$. This operator is unitary whenever
\begin{equation}\label{conditions}
\xi^+ =-\left(\xi^-\right)^* e^{i\,{\rm Im} \xi^z},\qquad|\xi^-|^2+1=e^{{\rm Re} \xi^z}.
\end{equation}
Note that the first condition above implies $|\xi^+| =|\xi^-|$.

Eq. \eqref{gauge transformation} leads to the following differential equations for  functions $\xi^\pm_t$, $\xi^z_t$:
\begin{align}\label{differential equation Gauss}
i \dot \xi^+_t =& \, B^-_t\left(\xi^+_t\right)^2-B_t^z \, \xi^+_t - B^+_t, \nonumber \\
i \dot \xi^z_t  = & \, 2 B^-_t\, \xi^+_t -B_t^z,  \nonumber \\
i \dot \xi^-_t   = &\,  - B^-_t\, \exp(  \xi^z_t ),
\end{align}
where $B^\pm_t\equiv(B^x_t\mp iB^y_t)/2$ (this definition implies $\B_t {\bf S}_{\rm tot} = B_t^+ S_{\rm tot}^+ + B_t^- S_{\rm tot}^- + B_t^z S_{\rm tot}^z$). The initial condition is $\xi^\pm_0=\xi^z_0=0$. In a somewhat different context, the system of equations  \eqref{differential equation Gauss} was derived in ref. \cite{ringel2013dynamical} following the lines of the early work \cite{wei1963lie}. It can be verified that these equations are consistent with the conditions~\eqref{conditions}.
Note that the first equation is a Riccati equation with a single variable and two others are trivially integrated when the solution of the first one is plugged in. These equations are somewhat simpler than the equivalent system \eqref{resolved}. In addition,  equations in \eqref{differential equation Gauss} have  non-singular right hand sides, which makes  clear that the solution exists for an arbitrary~$B_t$.

\end{appendix}

%
%

\bibliographystyle{spphys}       
\bibliography{C:/D/Work/QM/Bibs/1D,C:/D/Work/QM/Bibs/LZ_and_adiabaticity,C:/D/Work/QM/Bibs/scars,C:/D/Work/QM/Bibs/Floquet,C:/D/Work/QM/Bibs/dynamically_integrable,C:/D/Work/QM/Bibs/integrability}

\end{document}